\documentclass[showpacs,preprintnumbers,amsmath,amssymb,superscriptaddress]{revtex4}
\usepackage{graphicx}
\usepackage{dcolumn}
\usepackage{bm}
\usepackage{color}
\graphicspath{{EPS_figs/}{PDF_figs/}}

\begin{document}

\title{Elastic monopoles and external torques in nematic liquid crystal colloids}

\author{O. M. Tovkach}
\affiliation{Bogolyubov Institute for Theoretical Physics, NAS of Ukraine, Metrologichna 14-b, Kyiv 03680,Ukraine}

\author{S. B. Chernyshuk}
\affiliation{Institute of Physics, NAS of Ukraine, Prospekt Nauky 46, Kyiv 03650, Ukraine}

\author{B. I. Lev}
\affiliation{Bogolyubov Institute for Theoretical Physics, NAS of Ukraine, Metrologichna 14-b, Kyiv 03680,Ukraine}


\begin{abstract}
Up to now it is commonly believed that a colloidal particle suspended in a nematic liquid crystal never produces elastic monopoles because this violates the mechanical equilibrium condition. 
And the only way to obtain deformations of director field falling off with distance as $r^{-1}$ is to exert an external torque $\bm{\Gamma}_{\text{ext}}$ on the colloid \cite{de_Gennes}. 
In this paper we demonstrate that this statement is not quite correct and elastic monopoles, as well as dipoles and quadrupoles, can be induced without any external influence just by the particle itself. 
A behavior of a spherical colloidal particle with asymmetric anchoring strength distribution is considered theoretically. 
It is demonstrated that such a particle when suspended in a nematic host can produce director deformations decreasing as $r^{-1}$, i.e. elastic monopoles, by itself without any external influence.
\end{abstract}

\maketitle

Nematic liquid crystal colloids have attracted significant interest during the last decades. Particles, suspended in a liquid crystal host, break its continuous symmetry and cause director field distortions which may be accompanied by topological defects \cite{Poulin_1998, Stark_1999, Lubensky_1998}. The distortions, in turn, give rise to a new class of interactions that do not occur in usual colloids. These long-range anisotropic interactions result in different structures such as linear \cite{Poulin_1998, Poulin_1997} and inclined \cite{Poulin_1998, Smalyukh_2005, Kuzmin_2005, Kotar_2006} chains. Particles at a nematic-air interface as well as quasi two-dimensional colloids in thin nematic cells form a rich variety of 2D crystals \cite{Nazarenko_2001, Smalyukh_2004, Musevic_2006, Skarabot_2007, Skarabot_2008, Ognysta_2007}. 
Recently authors of \cite{nych1} observed 3D colloidal crystal structures in the bulk nematic liquid crystal.

Basis of theoretical description of these phenomena were outlined in \cite{Lubensky_1998, Ramaswamy_1996, Lev_1999, Lev_2002}. Their main idea is rooted in the fact that far from the particle director deviations $\delta\textbf{n}$ from its ground state $\textbf{n}_{0}$ are small. If we choose a coordinate system in such a way that $\textbf{n}_{0}=(0, 0, 1)$, the director field can be written as $\textbf{n}(\textbf{r}) \approx (n_{x}, n_{y}, 1)$, where $|n_{x}|, |n_{y}| \ll 1$. Then we are allowed to expand $n_{\mu}$, $\mu=\{x, y\}$, in multipoles
\begin{equation}\label{Multipoles_expansion}
n_{\mu} (\textbf{r}) = \frac{q_{\mu}}{r} + \frac{p_{\mu}^{\alpha} r_{\alpha}}{r^3} + \frac{Q_{\mu}^{\alpha\beta} r_{\alpha} r_{\beta}}{r^5} +...,
\end{equation}
where $\alpha$ and $\beta$ take values $x,y,z$ and summation over repeated greek indices is assumed. Coefficients $q_{\mu}$, $p_{\mu}^{\alpha}$, $Q_{\mu}^{\alpha\beta}$ are called elastic monopoles (charges), dipoles and quadrupoles, respectively. As it follows from \eqref{Multipoles_expansion}, director deviations $n_{x}$ and $n_{y}$ have a long-range nature. This means that the deformations caused by different particles can overlap even if the particles are located far from each other. Because of this the system cannot minimize its energy 	by minimizing all the deformations separately. They must be treated consistently. In practice the overlapping manifests itself in the fact that a colloidal particle "feels" the presence of the other particles mediated by a nematic host, i.e in the appearance of the effective long-range elastic interactions between colloidal particles. 

These elastic long-range interactions in bulk nematic colloids are determined completely by the coefficients $q_{\mu}$, $p_{\mu}^{\alpha}$, $Q_{\mu}^{\alpha\beta}$. In the case of strong anchoring they must be found from asymptotics of the solutions of nonlinear equations describing $\textbf{n}(\textbf{r})$ in the vicinity of the particle. But when the anchoring is weak $\delta\textbf{n}$ is small and consequently expansion \eqref{Multipoles_expansion} is valid everywhere outside the particle. Under these circumstances the multipole coefficients are determined by the symmetry of the particle surface \cite{Lev_2002}. For instance, expansion \eqref{Multipoles_expansion} always contains at least two quadrupole terms. Dipoles appear as a result of broken mirror symmetry \cite{Lev_2002}. But
up to now it is commonly believed that a colloidal particle itself, despite its symmetry, never produces elastic monopoles because this violates the mechanical equilibrium condition. And the only way to obtain deformations of director field falling off as $r^{-1}$ is to exert an external torque $\bm{\Gamma}_{\text{ext}}$ on the colloid \cite{de_Gennes}.

In this paper we demonstrate that this statement is not quite correct and elastic monopoles, as well as dipoles and quadrupoles, can be induced without any external influence just by the particle itself. We were interested in equilibrium orientations of a sphere with asymmetric distribution of the anchoring strength on its surface. It appeared that there are such equilibrium states in which elastic monopoles exist even if $\bm{\Gamma}_{\text{ext}}=0$.

It is well known that nematics, unlike isotropic liquids, transmit torques. As it was shown in \cite{de_Gennes},  torque $\bm{\Gamma}$, acting on NLC, may be written in the following form
\begin{equation}\label{Torque}
\bm{\Gamma} = \left[ \textbf{n} \times \frac{\delta F}{\delta \textbf{n}} \right],
\end{equation} 
where $F$ is the nematic free energy. Since the director deformations have energy
\begin{equation}\label{F_bulk}
F_{\text{def}} = \frac{K}{2} \int dV \left[ (\nabla \cdot \textbf{n} )^2 + (\nabla \times \textbf{n} )^2 \right], 
\end{equation}
they are coupled with some torque $\bm{\Gamma}_{\text{def}}$. But only monopoles make a nonzero contribution to $\bm{\Gamma}_{\text{def}}$
\begin{equation}\label{Monopole_dG}
\bm{\Gamma}_{\text{def}} = \left[ \textbf{n} \times \frac{\delta F_{\text{def}}}{\delta \textbf{n}} \right] = 4\pi K \textbf{q}^{T}
\end{equation}
where $\textbf{q}^{T} = (q_{y}, q_{x}, 0)$ and $\Gamma_{z}^{\text{def}}=0$ since a rotation around $\textbf{n}_0$ does not alter $F_{\text{def}}$. Deformations decreasing faster then $r^{-1}$ are not related with any torque. In turn, $\bm{\Gamma}_{\text{def}}$ can be treated as the torque we need to exert on a nematic to induce elastic monopoles $q_x$ and $q_y$ in there. Now let us assume that we have a particle immersed in some bulk sample of NLC and there are no external torques exerted on it, $\bm{\Gamma}_{\text{ext}}=0$. If there exist elastic monopoles  the particle will "feel" torque $-\bm{\Gamma}_{\text{def}}$ and, under these circumstances ($\bm{\Gamma}_{\text{ext}}=0$), will constantly rotate. Obviously this is not a physical situation. Therefore, we ought to state that the only source of elastic monopoles is the external torque exerted on the particle, $\bm{\Gamma}_{\text{ext}}=-\bm{\Gamma}_{\text{def}}$.  

But the point is that the energy of the colloidal system is not exhausted just by deformations. It contains the energy of the nematic-particle's surface interaction as well. This energy can be written in Rapini-Papoular form 
\begin{equation}
F_{\text{surface}} =\oint dS \,\, W(\textbf{s}) \big[ \bm{\nu}(\textbf{s}) \cdot \textbf{n}(\textbf{s}) \big]^2,
\end{equation}
where $W(\textbf{s})$ is the anchoring strength. As it was noted above in the weak anchoring case $\textbf{n} = \textbf{n}_0 + \delta \textbf{n}$, $\delta \textbf{n} \ll 1$ everywhere, and the surface energy gives rise to torque $\bm{\Gamma}_{\text{surface}}$
\begin{equation}\label{Torque_surface}
\bm{\Gamma}_{\text{surface}} = \left[ \textbf{n} \times \frac{\delta F_{\text{surface}}}{\delta \textbf{n}} \right] \approx 2 \oint dS\,W\left( \bm{\nu} \cdot \textbf{n}_{0} \right) \left[ \textbf{n}_{0} \times \bm{\nu} \right]
\end{equation}
If the particle has broken "horizontal" (i.e. perpendicular to the $\textbf{n}_0$) and at least one of "vertical" symmetry planes integrals \eqref{Torque_surface} can be nonvanishing. In the equilibrium the total torque acting on the system : particle + LC has to be zero  
\begin{equation}\label{Torque_balance}
\bm{\Gamma}_{\text{total}}= \bm{\Gamma}_{\text{ext}}+\bm{\Gamma}_{\text{def}}+\bm{\Gamma}_{\text{surface}}=0
\end{equation}
Hence, in the general case $q_{\mu}$ are produced by both external torque and particle itself
\begin{equation}\label{Monopole_Lev}
\textbf{q}^{T} = -\frac{\bm{\Gamma}_{\text{ext}}+ \bm{\Gamma}_{\text{surface}} }{4 \pi K}.
\end{equation}
We can also look at this issue from another viewpoint. In terms of mathematics, expression \eqref{Monopole_dG} is obtained from the divergence theorem. Indeed, volume integral $\frac{\delta F}{\delta \textbf{n}}$ may be transformed into some integral over a closed surface $\Sigma$. This implies that torques acting on the nematic bulk must be balanced by surface torques \cite{de_Gennes}. When we deal with a bulk nematic $\Sigma$ can be chosen at $r \to \infty $ and we come to \eqref{Monopole_dG}, i.e. torques associated with monopoles can be balanced only by external agents. But in a colloidal system we have slightly different situation. Besides $\Sigma$ there is the particle surface. And this real surface cannot be ignored in the divergence theorem and leads us to the expression \eqref{Monopole_Lev}. This fact is a simple illustration of the difference between electrostatics and nematostatics. If the electric charge (monopole) is a real physical point object, the elastic monopole is to a certain extent 
artificial object. The multipole expansion in nematostatics is just a way to describe deformations of the director field via point source. Although in fact they are produced by real particle surface.

\begin{figure}
\begin{center}
\includegraphics[width=4.5cm]{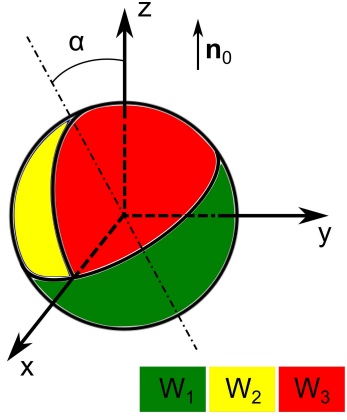}
\end{center}
\caption{Colloidal particle producing an elastic monopole. "Lower" hemisphere has an anchoring constant $W_1$. "Upper" hemisphere is divided into two equal parts with anchoring constants $W_2$ and $W_3$. The particle plane of symmetry coincides with $yz$-plane. The only parameter describing the particle orientation is the angle of rotation around $x$-axis denoted by $\alpha$.}\label{Sphere}
\end{figure}

Below we will try to clarify our statement on a concrete example. Let us consider a spherical particle suspended in a nematic liquid crystal. We divide its surface into three parts with different anchoring constants $W_{1}$, $W_{2}$ and $W_{3}$ (see Fig.~\ref{Sphere}). Such a particle has only one plane of symmetry. In the equilibrium state this plane has to coincide with one of the nematic's planes of symmetry. It can be either vertical, i.e. coinciding with coordinate $yz$-plane, or horizontal, i.e. coinciding with coordinate $xy$-plane. The former is the case we are interested in. Thus, due to the symmetry of the particle its orientation is determined only by angle $\alpha$ (Fig.~\ref{Sphere}).
Now we want to examine which one of equations \eqref{Monopole_Lev} and \eqref{Monopole_dG} with $\bm{\Gamma}_{\text{def}} =-\bm{\Gamma}_{\text{ext}}$ is correct. To do that let us assume that the anchoring is weak: $W_{k}a \ll K$, $a$ is the particle radius and $k = \{1,2,3 \}$.Then the free energy of the system under investigation can be written as
\begin{equation}\label{FE_example}
F = F_{\text{def}} + F_{\text{surface}} = \frac{K}{2} \int dV (\nabla n_{\mu} \cdot \nabla n_{\mu}) + \oint dS\, W(\textbf{s}) \nu_{z}^{2}(\textbf{s}) + 2 \oint dS\, W(\textbf{s}) \nu_{z}(\textbf{s})\nu_{\mu}(\textbf{s})n_{\mu}(\textbf{s}),
\end{equation}
where the terms like $W n_{\mu} n_{\mu^{\prime}}$ are neglected because of their smallness. Director deviations $n_{\mu}$ everywhere outside the particle are described by the following expressions
\begin{equation}
n_{\mu} (\textbf{r}) = \frac{q_{\mu}}{r} + \frac{p_{\mu}^{\alpha} r_{\alpha}}{r^3},
\end{equation}
where we have omitted the quadrupole terms. This does not alter our results qualitatively, but allows us to simplify calculations. Indeed, as was mentioned above, the director field can be written as $\textbf{n}(\textbf{r}) = (n_{x}, n_{y}, 1)$. This is the so-called harmonic approximation. To find the leading anharmonic corrections we should set $\textbf{n}(\textbf{r}) = (n_{x}, n_{y}, \sqrt{1-n_{\bot}^{2}}) \approx (n_{x}, n_{y}, 1-\frac{1}{2}n_{\bot}^{2})$. Then $n_{\mu}$ may be found from the following equations
\begin{equation}
\Delta n_{\mu} + \frac{1}{2} n_{\mu} \Delta n_{\bot}^{2} = 0,
\end{equation} 
where $n_{\bot}^{2}=n_{\mu}n_{\mu}$. Within the harmonic approximation they are solutions to the Laplace equations $\Delta n_{\mu} = 0$. Hence it follows that if the leading harmonic term in $n_{\mu}$ decreases as $r^{-n}$ then the first anharmonic correction will fall off as $r^{-3n}$ \cite{Lubensky_1998}. Therefore, when the monopole exists the quadrupolar terms can be ignored within the harmonic approximation.

Integrating \eqref{FE_example} we obtain the energy of the system as a function of the particle orientation and the director deformations, $F(\alpha, q_{x},q_{y}, p_{x}^{x}, p_{x}^{y}, p_{x}^{z}, p_{y}^{x}, p_{y}^{y}, p_{y}^{z}) = F_{\text{bulk}}(\alpha, q_{x},q_{y}, p_{x}^{x}, p_{x}^{y}, p_{x}^{z}, p_{y}^{x}, p_{y}^{y}, p_{y}^{z}) + F_{\text{surface}}(\alpha, q_{x},q_{y}, p_{x}^{x}, p_{x}^{y}, p_{x}^{z}, p_{y}^{x}, p_{y}^{y}, p_{y}^{z})$, where 
\begin{equation}
F_{\text{bulk}}(\alpha, q_{x},q_{y}, p_{x}^{x}, p_{x}^{y}, p_{x}^{z}, p_{y}^{x}, p_{y}^{y}, p_{y}^{z}) = \frac{2 \pi K}{a} (q_{x}^{2} + q_{y}^2) + \frac{4 \pi K}{3a^{3}}({p_{x}^{x}}^{2} +{p_{x}^{y}}^{2} +{p_{x}^{z}}^{2} +{p_{y}^{x}}^{2} +{p_{y}^{y}}^{2} +{p_{y}^{z}}^{2})
\end{equation}
and
\begin{multline}
F_{\text{surface}}(\alpha, q_{x},q_{y}, p_{x}^{x}, p_{x}^{y}, p_{x}^{z}, p_{y}^{x}, p_{y}^{y}, p_{y}^{z}) = \frac{\pi a^{2}}{3}(2W_1 +W_2 +W_3) +\frac{2a^2}{3}(W_3 -W_2)\sin 2\alpha\\
+\frac{\pi}{4} \left[ (W_2+W_3-2W_1)(p_{y}^{y}\cos^3\alpha -p_{y}^{z}\sin^3\alpha +p_{x}^{x}\cos\alpha) +(W_3-W_2)(p_{y}^{y}\sin^3\alpha +p_{y}^{z}\cos^3\alpha +p_{x}^{x}\sin\alpha) \right]\\
+\frac{4a}{3} (W_3 -W2)q_{y}  \cos 2\alpha.
\end{multline}
The mechanical equilibrium condition requires that the energy of the system $F(\alpha, q_{x},q_{y}, p_{x}^{x}, p_{x}^{y}, p_{x}^{z}, p_{y}^{x}, p_{y}^{y}, p_{y}^{z})$ should be minimal (this minimum can be global as well as local) $\frac{\partial F}{\partial \alpha}=\frac{\partial F}{\partial q_{x}}=...=0$ :
\begin{subequations}
\begin{multline}\label{a}  
\frac{4a^2}{3}(W_3 -W_2)\cos 2\alpha -\frac{8a}{3} (W_3 -W2)q_{y} \sin 2\alpha\\
+\frac{\pi}{4} (2W_1-W_2-W_3)(3 p_{y}^{y}\cos^2\alpha\sin\alpha +3 p_{y}^{z}\sin^2\alpha\cos\alpha +p_{x}^{x}\sin\alpha)\\
 +\frac{\pi}{4}(W_3-W2)(3p_{y}^{y}\sin^2\alpha\cos\alpha -3p_{y}^{z}\cos^2\alpha\sin\alpha +p_{x}^{x}\cos\alpha) = 0
\end{multline}
\begin{align}
&\frac{4\pi K}{a} q_{x} =0\label{b}\\
&\frac{4\pi K}{a} q_{y} +\frac{4a}{3} (W_3 -W_2)\cos 2\alpha =0\label{c}\\
&\frac{8\pi K}{3a^3} p_{x}^{x} +\frac{\pi}{4} \left[ (W_2+W_3-2W_1)\cos\alpha +(W_3-W_2)\sin\alpha  \right] =0\label{d}\\
&\frac{8\pi K}{3a^3} p_{x}^{y} =0,\qquad \frac{8\pi K}{3a^3} p_{x}^{z} =0,\qquad \frac{8\pi K}{3a^3} p_{y}^{x} =0 \label{e}\\
&\frac{8\pi K}{3a^3} p_{y}^{y} +\frac{\pi}{4} \left[ (W_2+W_3-2W_1)\cos^3\alpha +(W_3-W_2)\sin^3\alpha  \right] =0\label{h}\\
&\frac{8\pi K}{3a^3} p_{y}^{z} +\frac{\pi}{4} \left[ (2W_1 -W_2-W_3)\sin^3\alpha +(W_3-W_2)\cos^3\alpha  \right] =0\label{i}
\end{align}
\end{subequations}
Substituting equations \eqref{b} - \eqref{i} into equation \eqref{a}, one can easily find that $\alpha$ obeys the following 
\begin{equation}\label{eq_alpha}
A \cos 2\alpha + B \sin 2\alpha + C \sin 4\alpha = 0,
\end{equation}
where $A = 6\pi (W_2 -W_3) (-512K +9\pi a (W_2 + W_3 - 2W_1))$, $B = 108 \pi^{2} a (W_1 - W_2)(W_1 - W_3)$ and $C = a (1024(W_2 -W_3)^{2} +81 \pi^{2} (2W_1^{2} +W_2^2 +W_3^2 -2W_1(W_2 +W_3)))$. 

For further analysis it is convenient to rewrite \eqref{eq_alpha}, using the substitution $\tan \alpha =x$, then
\begin{equation}\label{eq_x}
A x^4 -(2B -4C)x^3 -(2B-4C)x -A=0.
\end{equation} 
It is easy to see that the left side of \eqref{eq_x} can be treated as a  a continuous function of $x$. The function values have opposite signs at the ends of the intervals $(-\infty, 0] \,\, \text{and} \,\, [0, +\infty)$. Hence, equation \eqref{eq_x} always has real roots. On the other hand from equation \eqref{c} it follows that $q_{y}$ vanishes only if $\alpha = \dfrac{\pi}{4} + \dfrac{\pi n}{2},\,\, n \in \mathbb{Z}$. But these $\alpha$ are not solutions of  equation \eqref{eq_alpha}. It means that our system has such equilibrium states in which the elastic charge exists without any external torque.

\begin{figure}
\begin{center}
\includegraphics[width=.72\textwidth]{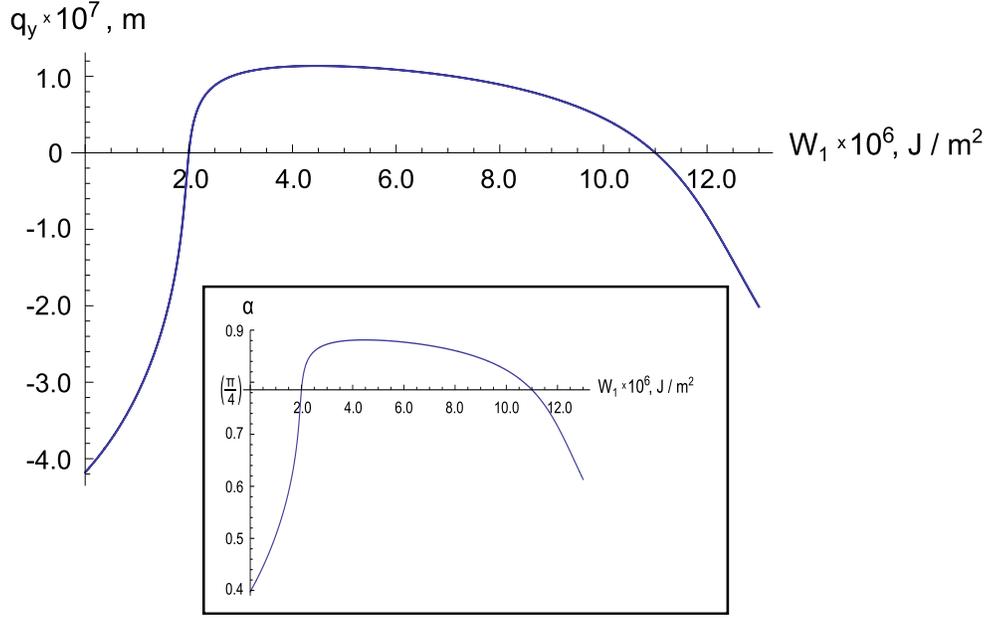}
\end{center}
\caption{Dependence of elastic monopole $q_y$ on anchoring strength $W_1$. Here $W_2 = 11 \cdot 10^{-6}\,J/m^2$, $W_3 = 2 \cdot 10^{-6}\,J/m^2$, $K = 10\,pN$, $a = 2.5\,\mu m$. Inset shows the particle orientation as a function of $W_1$.}\label{Monopole_plot}
\end{figure}

\begin{figure}
\begin{center}
\includegraphics[width=11.5cm]{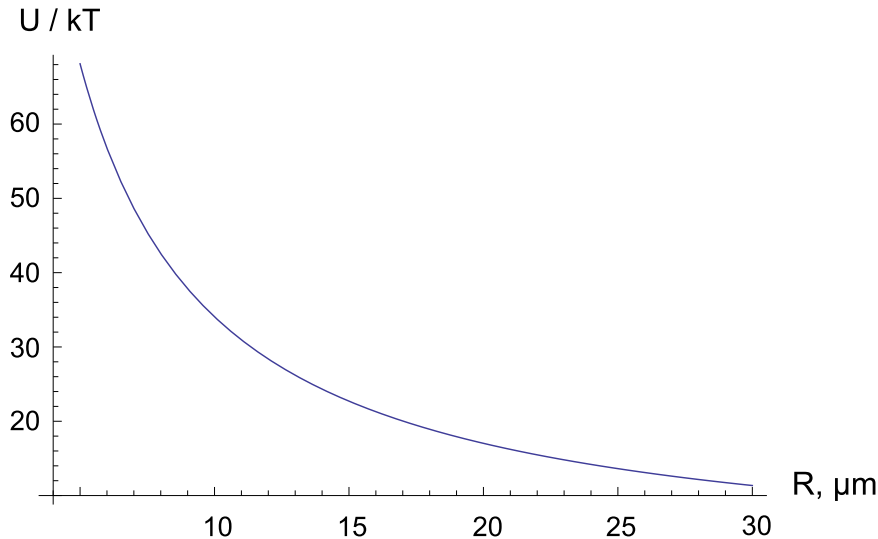}
\end{center}
\caption{Energy of the monopole-monopole repulsion in $kT$ units as a function of interparticle distance $R$. Here $q_{y} = -q_{y}^{\prime} = 1 \cdot 10^{-7}\, m$ and $U = 4\pi K qq^{\prime} / R$.}\label{Energy_plot}
\end{figure}

Expression \eqref{Torque_surface} implies that the particle with at least two orthogonal planes of symmetry cannot be a source of elastic monopoles. A simple illustration of this fact is given in Fig.\ref{Monopole_plot} which shows the dependence of $q_y$ and the particle orientation on the anchoring strength $W_1$. We clearly see that $q_y=0$ when an additional symmetry plane appears, that is, when $W_1=W_2$ or $W_1=W_3$. Besides, as it follows from Fig.\ref{Energy_plot} typical values of $q_y$ are of order of $10^{-7}\, m$. This, in turn, means that the elastic monopoles can be observable even if the anchoring is weak. Indeed, the energy of repulsion between two identical monopoles $q_{y} = -q_{y}^{\prime} = 1 \cdot 10^{-7}\, m$ separated by $R = 20\, \mu m$ is of order of $10\,kT$, more precisely $U_{\text{qq}} = -4\pi K q_{y}q_{y}^{\prime} / R \approx 17\,kT$ (see Fig.\ref{Energy_plot}). Note that the elastic monopoles of the opposite signs repel each other while monopoles of the same signs attract. 

To conclude we have found on the concrete example that the surface of the particle can exert nonzero torque on a nematic liquid crystal even without any external influences. In this case asymmetrical anchoring distribution $W(s)$ plays the role of some external field. The total torque of the particle remains zero while the elastic charge can be effectively produced. Thus, we expect that asymmetric particles with strong anchoring will produce monopoles as well.
Therefore elastic monopoles as well as dipoles and quadrupoles can exist in the nematic bulk without any external torques.

\end{document}